\documentclass[onecollarge]{svjour2}
\usepackage{times}
\usepackage{mathptmx}

\usepackage{epsfig}
\usepackage{longtable}
\usepackage{subfigure}
\usepackage{amssymb}
\usepackage{graphicx}
\begin{document}

\bibliographystyle{plain}

\newcommand{\fullversion}[1]{}
\newcommand{\extabs}[1]{#1}
\newcommand{\swallow}[1]{}

\title{Call admission control algorithm for pre-stored VBR video streams
%\extabs{\\(Extended Abstract)}
}

\author{C. Tryfonas \and D. Papamichail \and A. Mehler \and S.~S. Skiena}

\institute{C. Tryfonas \at Kazeon Systems, Inc., 1161 San Antonio Road , Mountain View, CA 94043, USA
\email{tryfonas@kazeon.com}
\and 
D. Papamichail 
\at Computer Science Department,
University of Miami,
Coral Gables, FL 33146, USA
\email{dimitris@cs.sunysb.edu}
\and A. Mehler \and S.~S. Skiena \at
Computer Science Department,
SUNY at Stony Brook,
Stony Brook, NY 11794, USA
\email{\{mehler$|$skiena\}@cs.sunysb.edu}
}

\maketitle

\abstract{
We examine the problem of accepting a new request for a pre-stored VBR video stream that
has been smoothed using any of the smoothing algorithms found in the literature.
The output of these algorithms is a piecewise constant-rate schedule for a Variable Bit-Rate (VBR) stream. 
The schedule guarantees that the decoder buffer does not overflow or underflow.
The problem addressed in this paper is the determination of the minimal time displacement of each 
new requested VBR stream so that it can be accomodated by the network and/or the video server without
overbooking the committed traffic. We prove that 
this call-admission control problem for multiple requested VBR streams is NP-complete and inapproximable within
a constant factor, by reducing it 
from the {\sc Vertex Color} problem. 
We also present a deterministic morphology-sensitive algorithm that calculates the minimal 
time displacement of a VBR stream request. The complexity of the proposed algorithm make it suitable for
real-time determination of the time displacement parameter during the call admission phase.
}


\section{Introduction}
\label{sec:introduction}

% generic paragraph about video services

A significant portion of the forecasted network traffic is expected
to be multimedia (e.g. voice and video)
traffic. New services such as video-on-demand (VoD) and TV broadcasting
are currently under massive deployment.
One of the salient characteristics of video traffic is that it usually exhibits 
high variability in its bandwidth demands in different time scales.
The need to better understand the bandwidth demands of video streams
is essential for proper resource provisioning of both the network resources and
the resources of the video servers when stored video is transported. Proper
resource dimensioning has direct correlation with the quality of the recovered
video on the decoder and, therefore, a variety of techniques have been proposed 
in the past.

% Statistical CAC approaches

Significant work has been done in the literature
in the area of statistical modeling
of video traffic for resource provisioning purposes,
so that it can be effectively transported over
packet-switched networks~\cite{casilari_itc99,elsayed95,elwalid_infocom93,guerin_jsac91,kesidis94}.
In most cases, the objective of these efforts is to build a general model that can
be used for resource dimensioning for all the video traffic
transported over the network. In some cases, the long-range
dependence (LRD) characteristic of video traffic is
exploited to create a model of the traffic
source~\cite{casilari_itc99,gao-multifractal,li_2000}.
These methods, in general,
characterize the traffic source based on its statistical properties,
and provide value when the video stream is not known a-priori.
However, when dealing with pre-stored video, the resource dimensioning
process can be made deterministic and any statistical technique is of limited
value since it does not capture the exact dynamics of the video 
stream in the time domain.

% smoothing techniques - they are used to smooth the video traffic given 
% resource contraints (mainly decoder buffer) 

In video applications that transport stored video over a packet-switched
network, the resource provisioning process can take advantage of the fact
that video streams can be pre-processed offline. During the pre-processing of a video
stream, a transmission schedule is typically computed to minimize its rate variability and,
therefore, facilitate the resource provisioning and the call admission control process.
The reduction in rate variability is done by work-ahead smoothing, i.e. sending more data to the 
receiver with respect to its playback time. Significant work can be found in the literature 
in the area of work-ahead
video smoothing~\cite{feng_book,jiang_infocom98,mcmanus_jsac96,salehi_sigmetrics96}. The general
idea behind most of these algorithms is to maximize the
time intervals (rate segments) at which a transmission rate for the video stream
is used without causing under/overflow of the receiver buffer.
The algorithms differ in the selection of the starting point of these rate segments. 
The output of these algorithms is a piecewise constant-rate schedule for the smoothed video stream. 
The schedule guarantees that the decoder buffer does not overflow or underflow.
Due to the fact that computing the smoothing schedule is not a trivial process and cannot
be performed online, the pre-computed smoothed schedules of a video file for various decoder profiles 
can be stored along with the file itself in the video servers,
so that they can be used at the time of the corresponding video request to guarantee 
a deterministic quality at the decoder.

% problem definition

The problem addressed by this paper is that of accepting a new request for a pre-stored VBR video stream 
that has been pre-smoothed using any of the smoothing algorithms. Since the request can come at 
any particular point in time, the problem is related to the accomodation of the new request provided that the 
envelope of the dynamics of the committed traffic, and therefore,
the envelope of the available bandwidth in the channel, does not introduce overallocation at any time
interval (see Fig. 1). The goal is to displace the pre-computed smoothed schedule of the new request 
into the future to avoid overallocation. More specifically, we want to find the minimum time displacement
of the new schedule so that the channel can accomodate the new request. The problem
described is an optimization problem that can be extended in several ways. For example,
given a set of requests, find the displacement points of the associated schedules so that the
overall schedule is the smoothest.

% our approach

In this paper we present two algorithms that solve the problem of computing the
minimum dispacement of a new request, also referred as the {\sc two Stream Scheduling} problem ({\em 2-SS}): 
(i) a simple algorithm with $O(n^2logn)$ 
complexity, and (ii) a morphology-sensitive algorithm with lower computational complexity.
The morphology-sensitive algorithm makes specific observations about the smoothed schedule so that certain peaks
can be skipped by the algorithm to speed-up the final calculation considerably, depending on the input.
%Both the correctness and complexity analysis of the proposed algorithms are presented.
Then we present a lower bound on the complexity of the {\em 2-SS} problem, which
is shown to belong in the {\it 3SUM}-hard problem group.
We also demonstrate that the problem of computing the minimum displacement of multiple new requests
(also referred to as the {\sc Multiple Stream Scheduling} problem or {\em m-SS}) is NP-complete, and cannot be 
polynomially approximated within a constant factor.
This is proven by reducing the {\sc String Pack} problem to {\em m-SS}. {\sc String Pack} was
introduced and shown NP-complete in \cite{NIC05}. To obtain the approximability results,
we further reduce {\sc Vertex Color}\cite{SKI98} to {\sc String Pack}. This reduction yields
new hardness of approximability bounds for {\sc String Pack},
thus improving previous results.

% presentation of the paper skeleton

The rest of this paper is organized as follows:
In Section~\ref{sec:scheduling_2_streams},
we present the formal definition of the problem for call admission of two VBR streams (or 
equivalently the admission control of a new request over the envelope of available bandwidth in a channel).
We also propose two algorithms that are efficient for the {\em 2-SS} problem.
In Section~\ref{sec:scheduling_M_streams}, we extend the problem to multiple streams. 
We prove that the problem of admitting multiple streams is NP-complete by reducing
the {\sc String Pack} problem to it. Finally, in Section~\ref{sec:conclusions}
we conclude the paper with a summary of this work.

\section{Admission Control of a new request}
\label{sec:scheduling_2_streams}

\subsection{Formal definition}

The input of the {\sc Two Stream Scheduling} ({\em 2-SS}) problem is two ortholinear traffic 
envelopes (streams) $S_1$ and $S_2$ of total length $L_1$ and $L_2$ respectively and the channel 
bandwidth $B$. A stream envelope $S_k$ can be described 
by an ordered set of triplets $r_i = ({h_k}_i, {s_k}_i, {e_k}_i), i = 1\cdots n$, with ${h_k}_i$
being the height value (bandwidth demand of video) and ${s_k}_i$ and ${e_k}_i$ the starting and ending
time points of the $i$th peak respectively, of a total of $n$
non-overlapping peaks in the stream. Let ${l_k}_i = {e_k}_i - {s_k}_i$ be the length of the $i$th peak. 
For the {\em 2-SS} problem, $S_1$ consists of $n$ such triplets and $S_2$ of $m = O(n)$ triplets. 

\begin{figure}%[b]
\centering
%\label{fig:streams}
\epsfig{file=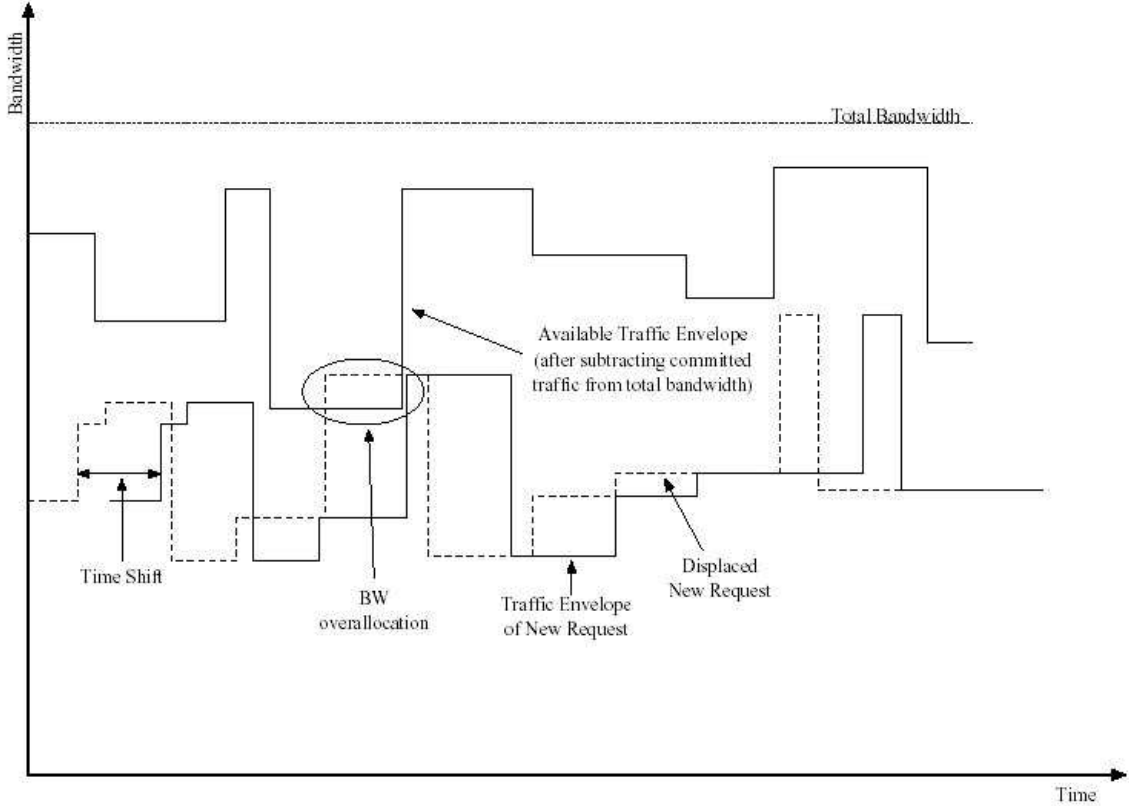, width=1.00\textwidth}
\caption{Example of accomodating a new video call request through time displacement.\label{fig:streams}}
\end{figure}

We consider $S_1$ being requested and transmitted at time point $0$,
so being fixed at that position. This allows us to subtract its content allocation
from the total bandwidth, creating a reverse envelope, as in Fig. \ref{fig:streams}.
Basically, stream $S_1$ corresponds to the committed traffic.
The second stream can be displaced by a positive time interval $T$ to its right, resulting
in delayed transmission. We assume that the envelopes are rigid and none of
the peaks can be altered either in length or height. The order of the 
peaks is fixed.

\begin{figure}%[b]
\centering
\epsfig{file=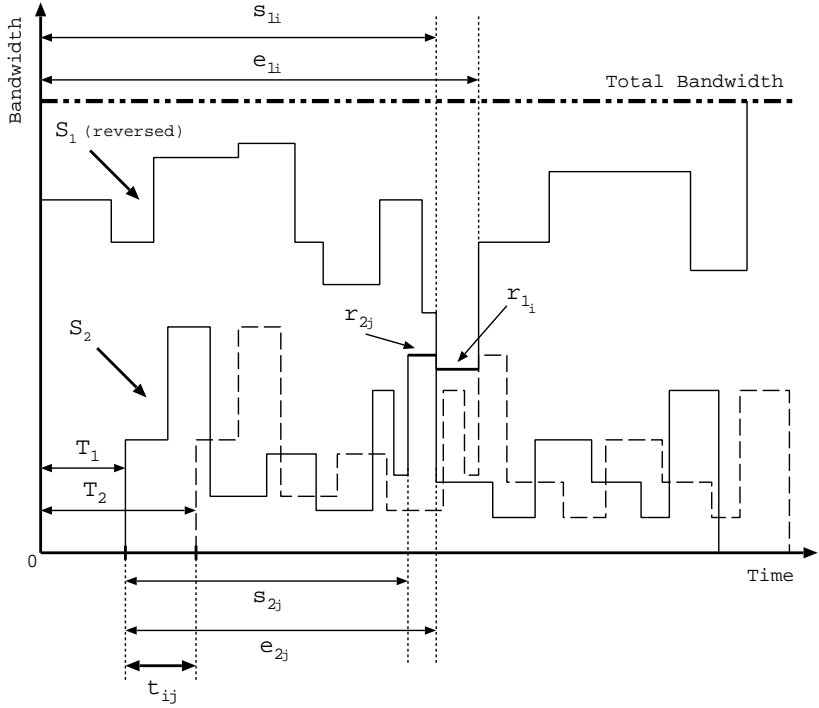, width=0.70\textwidth}
\caption{Intersection interval parameter display.\label{fig:intersection}}
\end{figure}

Let $S_2$ be displaced by $T$ time units. An intersection (time overlap) of 
the ${r_1}_i = ({h_1}_i, {s_1}_i, {e_1}_i)$ peak triplet from $S_1$ with the 
${r_2}_j = ({h_2}_j, {s_2}_j, {e_2}_j)$ peak triplet from 
$S_2$ occurs when ${l_1}_i + {l_2}_j > B$ and 
$\exists t \in [{s_1}_i, {e_1}_i]: t \in [{s_2}_j + T, {e_2}_j + T]$. The set of all
time points such that ${r_1}_i$ intersects ${r_2}_j$ defines
a time interval (referred to from now on as intersection interval) $t_{ij}$ of length 
${h_1}_i + {h_2}_j$, starting at
time point $T_1 = {s_1}_i - {e_2}_j$ and ending at $T_2 = {e_1}_i - {s_2}_j$. 
Intersection parameters are depicted
graphically in Fig. \ref{fig:intersection}. The second stream cannot be displaced by any
value corresponding to this intersection interval, or there will occur a bandwidth overallocation.

The output of the {\em 2-SS} algorithm will be the minimum displacement of $S_2$,
such that there is no bandwidth overallocation. The second stream can be shifted only by a displacement
that does not fall into any intersection interval $t_{ij}$, for $1<i<n$ and $1<j<m$. So, the output of the
algorithm could be described as the minimum displacement that does not fall into an intersection interval.

\subsection{A morphology sensitive algorithm}

In this section we describe an algorithm to solve the {\em 2-SS} overallocation problem.
The algorithm processes all segments, in order to calculate their
intersection interval. It could be the case though that many peaks will not be as high as
to intersect. By sorting the peaks by height (bandwidth demand), one can actually calculate the intersection
intervals only for the ones that actually intersect and not consider the rest.

Let $P$ be the number of peak pairs, where the first peak is selected from envelope $S_1$ and
the second from envelope $S_2$, that have sum of heights greater than the bandwidth $B$ and 
thus define an intersection interval. The algorithm then goes as follows: 

\begin{enumerate}
\item Sort the peak information (triplets) of both envelopes according to height.
\item Iterate through sorted peaks in $S_1$ and calculate their intersection interval
with all peaks from the sorted list of $S_2$ that cause bandwidth overallocation.
Stop when the height of the next peak in $S_1$ does not intersect the highest peak of $S_2$.
\item Sort all intervals according to their starting point.
\item Iterate through sorted intervals, merging them into an aggregate interval,
until an interval that does not intersect the aggregate interval is discovered, or
we run out of intervals.
\item Output the end point of the aggregate interval as the solution.
\end{enumerate}

\begin{figure}[htbp]
\centerline{\psfig{figure=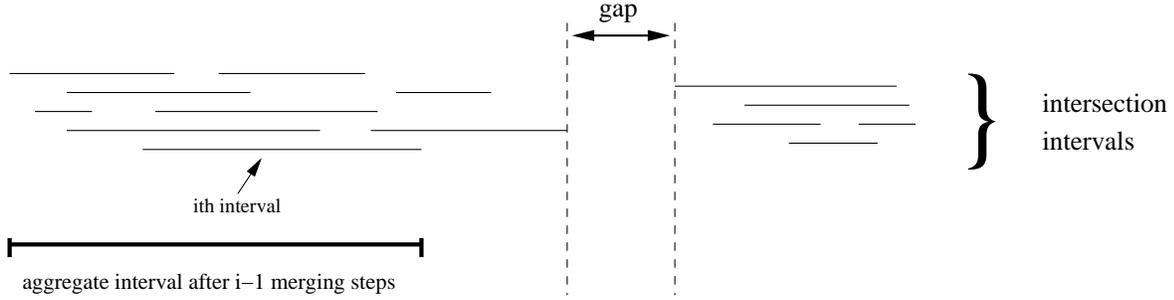,width=1.00\textwidth}}
\caption{
  Merging intersection intervals into an aggregate interval at the $i-1$ step. The start
  of the gap defines the $S_2$ stream displacement where no bandwidth overallocation occurs.
}
\label{fig:aggregate_interval}
\end{figure}

For the correctness of the algorithm we can argue that by iterating through all
intersecting peaks of both streams, we have discovered all possible time intervals where the
second stream cannot be shifted. The first ``gap'' between the aggregate interval
and the currently examined interval will provide the minimum displacement, since
any position in the aggregate interval defines a forbidden displacement, belonging
to some previously examined intersection interval. The start of the ``gap''
described above cannot belong to any interval, since, if such an interval existed,
its start would occur before the end of the aggregate interval and, as such, before
the currently examined interval, which means it would have already been included in the
aggregate interval. A visual representation of the procedure can be seen in Fig.
\ref{fig:aggregate_interval}.

Sorting the peaks by height takes $O(n\log(n))$ time, the iteration through sorted
peaks takes $O(P)$ time and sorting all intersection intervals takes $O(P logP)$ time.
The merging iteration takes at most $O(P)$ time. So the total complexity is $O(\left( P+n \right)logn)$.

It should be noted that in the worst case senario where the number of intersecting peaks 
between the two streams is $O(n^2)$, the assymptotic complexity becomes $O(n^2logn)$,
dominated by sorting the intersection intervals' starting points.
We can further improve the running time of this algorithm by excluding intersection interval
calculation for peak pairs that result in negative second stream displacement, although
such an optimization does not result in any assymptotic gain.

\subsection{{\em 2-SS} scheduling is {\em 3SUM} hard}

In this section we will prove that {\em 2-SS} is {\em 3SUM}-hard, a class of
problems introduced in \cite{GAJ95}. The {\em 3SUM} problem is to decide whether there exist
integers $a, b, c$ in a set of $n$ integers, such that $a+b+c=0$, which is currently 
considered to have complexity $\Theta(n^2)$.

The notion of {\em 3SUM}-hardness (or $n^2$-hardness) is formally introduced in \cite{BAR01,GAJ95},
the notation of which we follow. 
In brief, we will mention that a problem is considered {\em 3SUM}-hard if any instance of
the {\em 3SUM} problem can be reduced to some instance (with a comparable size) of the other 
problem in $o(n^2)$ time, where $n$ is the size of the input.

For our proof, we will need the following definition:
\begin{definition}
Given two problems $PR1$ and $PR2$ we say that $PR1$ is f(n)-solvable using $PR2$ if 
every instance of $PR1$ of size $n$ can be solved by using a constant number of 
instances of $PR2$ (of size $O(n)$) and $O(f(n))$ additional time. We denote this by
\begin{displaymath} 
PR1 \lll_{f(n)} PR2
\end{displaymath}
\end{definition}

To prove that {\em 2-SS} is {\em 3SUM}-hard, it will be sufficient to show
that another {\em 3SUM}-hard problem is $o(n^2)$-solvable using {\em 2-SS}.
For that purpose, we will use the following {\em 3SUM}-hard problem:

\noindent
{\bf Problem: }
$SCP$ (Segments Containing Points): Given a set $P$ of $n$ real 
numbers and a set $Q$ of $m=O(n)$ pairwise-disjoint intervals of real numbers,
is there a real number translation $u$ such that $P+u \subseteq Q$? $P+u$ here
indicates the set of intervals in $P$ translated by $u$.

$SCP$ was shown {\em 3SUM}-hard in \cite{BAR01}. We will now prove the following:

\begin{theorem}
$SCP \lll_{nlogn} 2-SS$
\end{theorem}
\begin{proof}
Given an instance of the $SCP$ problem, we construct two streams in the following way:
The $m$ intervals in set $Q$ and $n$ real numbers in set $P$ are sorted and Stream $S_1$ 
is constructed to have peaks of height $0$ on these intervals and peaks of 
height $1$ in between. The length $L_1$ of $S_1$ is
determined by the start of the first interval $s_1$ and end of the last interval $e_m$ in the sorted list and 
starts at time point 0, with every segment displaced in time by subtracting $s_1$ from each of
its coordinates.
Stream $S_2$ is constructed with peaks of length $\varepsilon$ with $\varepsilon \rightarrow 0$
of height 1 at locations defined by the sorted numbers of set $P$, with peaks of height $0$ in
the intervals in between. The length $L_2$ of $S_2$ is again determined from the smallest
and largest elements of $P$ ($p_1$ and $p_n$ respectively) and original displacement 
$T$ of $0$ is achieved by subtracting $p_1$ from all peak segment coordinates. 
We set the channel bandwidth $B=1$. The construct can be seen in Fig. \ref{fig:2SS_construct}.

\begin{figure}%[b]
\centering
\epsfig{file=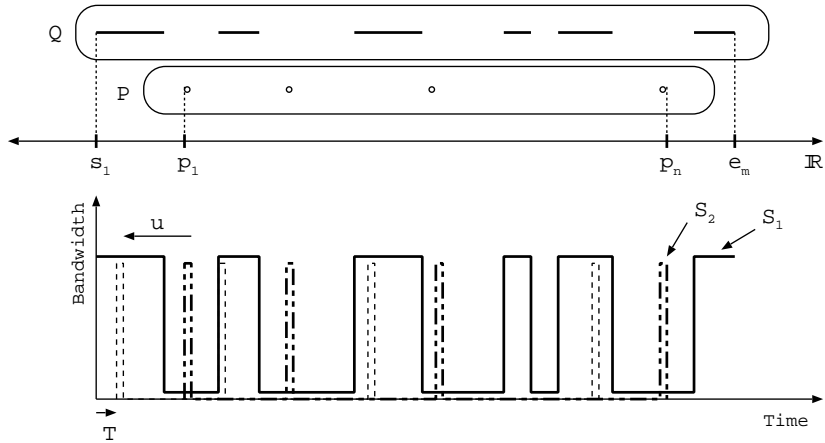, width=0.70\textwidth}
\caption{2SS construct. $S_1$ was translated in height for better viewing.\label{fig:2SS_construct}}
\end{figure}

We will now argue that the instance of $SCP$ has a solution if and only if the corresponding
instance of {\em 2-SS} has a displacement solution less than $L_1 - L_2$ (if $L_2 > L_1$ there
is no solution). From the construction it is 
obvious that a peak of $S_2$ of height $1$ can fit under a $0$ height peak of stream $S_1$ only if
the corresponding number in $P$ falls in the corresponding interval of $Q$. If for a certain 
displacement $T$ of $S_2$ we have $T < L_1 - L_2$ and there is no overallocation of bandwidth,
then all peaks of $S_2$ of height $1$ fit under $0$-height peaks of stream $S_1$, which would
imply that $\exists u = T + s_1- p_1: P+u \subseteq Q$. Also, by the same arguments, 
if there $\exists u: P+u \subseteq Q$, then $S_2$ displaced by $T = u + p_1 - s_1$ will result in
scheduling the two streams with no overallocation.

\end{proof}

Based on this result, we can conclude that our morphology-sensitive algorithm for scheduling two
streams is within a log factor from optimality.

\section{Scheduling multiple streams}
\label{sec:scheduling_M_streams}

We now extend the {\em 2-SS} problem to {\sc Multiple Stream Scheduling} ({\em m-SS}), where the input would consist
of multiple VBR streams that we want to schedule for transmission over a fixed bandwidth channel. 
Although we could set different objectives for optimization, we will select minimizing the 
displacement of the last stream being transmitted. For streams of the same size this is equivalent with
minimizing the 
total length of trasmission, starting from the time point of the first stream transmission and ending
when the last stream has been transferred over the channel.

\subsection{Multi-stream scheduling is NP-complete}

To demonstrate {\em m-SS} is NP-complete, we will reduce the {\sc String Pack} problem to it.
The {\sc String Pack} problem appeared in \cite{NIC05} and was proved hard by reduction from {\sc 3-Partition}.

\begin{figure}[htbp]
\centerline{\psfig{figure=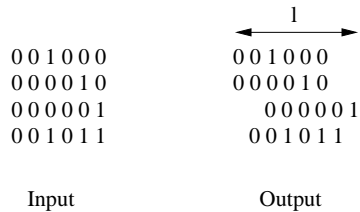, width=0.30\textwidth}}
\caption{
{\sc String Pack} example with $m=4, n=6 and l=8$
}
\label{fig:string_pack_example}
\end{figure}

The {\sc String Pack} is defined as follows: Given a set of $m$ strings of length $n$, over the binary alphabet
$\Sigma = \{0,1\}$, find a minimum length $l$ {\em packing} (alignment) of the strings, such that no column has more than
one `1'. An example of the input and the output of the problem are shown in Fig. \ref{fig:string_pack_example}.

\begin{figure}[htbp]
\centerline{\psfig{figure=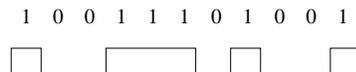, width=0.30\textwidth}}
\caption{
{String and equivalent stream transformation}
}
\label{fig:reduction_example}
\end{figure}

The reduction is straightforward. We will transform the input binary strings into streams with peaks of
height $1$ for each '1' encountered in the string and peaks of height $0$ for each '0' appearing in the
string, as shown in Fig.~\ref{fig:reduction_example}

We let the total available bandwidth $B=1$, such that no peaks from any stream can overlap. This
adheres to the requirement of the {\sc String Pack} problem not having any column with more than one '1'.
Since the input to the {\sc String Pack} problem is a set of strings with equal length $n$, minimizing
the total length of the outputed alignment is equivalent to minimizing the displacement of the last string.
Thus, the output of the {\em m-SS} on the transformed strings provides that exact minimum. So we have the following:

\begin{theorem}
{\sc String Pack} ${\leq}_p$ {\em 2-SS}
\end{theorem}
\begin{proof}
Given an instance of {\sc String Pack}, create a {\em 2-SS} instance by transforming the binary strings
to equivalent streams as described above. The minimum displacement of the last stream to be transmitted,
added to the length $n$ of the strings, provides the minimum length of the $m$ strings' packing.
\end{proof}

The result that {\em m-SS} is NP-complete follows from the observation that given a string packing, it
can be verified in time $O(mn)$ (thus polynomial in the input length) that it constitutes a valid solution, where
no column in the packing has more than one '1', and that the length of the packing is less than a 
specified length $k$, which would be an input of the decision version of the problem.

\subsection{{\sc Multiple Stream Scheduling} is polynomially inapproximable within a constant}

{\sc Vertex Color} is a well known problem\cite{SKI98}, defined as follows:
Given a graph $G=(V,E)$, color the vertices of $V$ with the minimum number of colors 
such that for each edge $(i,j) \in E$, vertices $i$ and $j$ have different colors. 

It has been shown that {\sc Vertex Color} is inapproximable within $|V|^{1-\epsilon}$ 
for any $\epsilon > 0$, unless ${\sc Zpp} = NP$
\cite{FEI98}. By reducing {\sc Vertex Color} to {\sc String Pack} and with the reduction
of the latter to {\em m-SS}, shown in the previous section,
we will demonstrate that any constant approximation of {\em m-SS} is NP-hard.

\subsubsection{Vertex Color}
We now show that {\sc VERTEX COLOR} reduces to {\sc
STRING PACK}; and that this reduction also yields a
polynomial approximation reduction.

Consider a graph $G=(V,E)$, and its vertex-edge
incidence matrix.  As a running example, we will use the graph given in
(Fig.~\ref{fig:examplegraph}) whose incidence matrix is shown below.

\[ 
\begin{array}{c}
\\ v_1 \\ v_2 \\ v_3 \\ v_4
\end{array}
\left(
\begin{array}{cccccc}
e_1 & e_2 & e_3 & e_4 & e_5 & e_6\\
0 & 0 & 1 & 0 & 0 & 0 \\
0 & 0 & 0 & 0 & 1 & 0 \\
0 & 0 & 0 & 0 & 0 & 1 \\
0 & 0 & 1 & 0 & 1 & 1 \\ \end{array}\right) \]

\begin{figure}[htbp]
\centerline{\psfig{figure=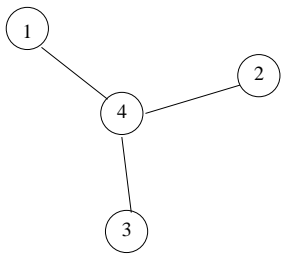,height=1.3in}}
\caption{
  A Graph on 4 vertices.
  }
\label{fig:examplegraph}
\end{figure}

It is clear that the graph can be colored with 2 colors.  $v_4$ gets one color, and $\{v_1, v_2, v_3\}$
get another color.  Also, the rows of the incidence matrix corresponding to a color group can all be
packed with no offset.  For example, putting together the rows for $\{v_1, v_2, v_3\}$ gives

\[ 
\begin{array}{c}
\\ v_1 \\ v_2 \\ v_3 \\ \mbox{Sum}
\end{array}
\left(
\begin{array}{cccccc}
e_1 & e_2 & e_3 & e_4 & e_5 & e_6\\
0 & 0 & 1 & 0 & 0 & 0 \\
0 & 0 & 0 & 0 & 1 & 0 \\
0 & 0 & 0 & 0 & 0 & 1 \\
0 & 0 & 1 & 0 & 1 & 1 \end{array}\right) \]

But if we try to pack strings from adjacent vertices, we will always get a collision (X); since
adjacent vertices have an edge in common.  For example

\[ 
\begin{array}{c}
\\ v_1 \\ v_4 \\ \mbox{Sum}
\end{array}
\left(
\begin{array}{cccccc}
e_1 & e_2 & e_3 & e_4 & e_5 & e_6\\
0 & 0 & 1 & 0 & 0 & 0 \\
0 & 0 & 1 & 0 & 1 & 1 \\
0 & 0 & X & 0 & 1 & 1 \end{array}\right) \]

So if we pack rows of the incidence matrix, the vertices the rows correspond to must all be
non-adjacent (i.e. can have the same color in a coloring).  But if we can color a graph with $c$ colors,
then we would be able to pack the rows of the incidence matrix into $c$ groups.  This is a good start,
but the {\sc String Pack} problem has no way of enforcing {\em groups}.  The strings are allowed to overlap
an arbitrary amount.  For instance, with the example matrix, it may give the following as a solution:

\[ 
\begin{array}{c}
v_1 \\ v_2 \\ v_3 \\ v_4
\end{array}
\left(
\begin{array}{cccccccc}
0 & 0 & 1 & 0 & 0 & 0 \\
0 & 0 & 0 & 0 & 1 & 0 \\
  &   & 0 & 0 & 0 & 0 & 0 & 1 \\
  & 0 & 0 & 1 & 0 & 1 & 1 \\ \end{array}\right) \]

In order to complete the reduction, we need to flank the incidence matrix with special strings that
will force any feasible solution to group the strings in the manner we desire.  That is, we want
strings to overlap completely, or not at all.  That way, if {\sc String Pack} gives a solution with $c$
groups, we know there must exist a $c$ coloring of $G$.

It turns out that we can not construct strings that overlap completely, or not at all.  We can make
strings that overlap completely, or by some small, bounded, amount.  This turns out to be sufficient.
Consider what overlaps we want to allow and dissallow.  In the following, an 'x' represents a region of
the flanking region, and the 0,1 are the rows of our incidence matrix.  The following 2 types of alignments
should be allowed by the flanking regions.

Complete overlap of incidence matrix rows.
\[ 
\left(
\begin{array}{cccccccccccccccccccc}
x & x & \ldots & x & 001000 & x & x & \ldots & x \\
x & x & \ldots & x & 000001 & x & x & \ldots & x \end{array}\right) \]

Zero overlap of incidence matrix rows.
\[ 
\left(
\begin{array}{ccccccccccccccccccccccccccccccccccc}
x & x & \ldots & x & 001000 & x & x & \ldots & x \\
  &   &        &   &        &   & x & x & \ldots & x & 000001 & x & x & \ldots & x \end{array}\right) \]

But we do not want to allow these following types of alignments, as they interfere with our 'grouping' of the
incidence matrix rows.

Interference of flanking region with incidence matrix rows.
\[ 
\left(
\begin{array}{ccccccccccccccccccccccccccccccccccc}
x & x & \ldots & x & 0 & 0 & 1 & 0 & 0 & 0 & x      & x & \ldots & x \\
  &   &        &   &   &   &   & x & x & \ldots & x & 0 & 0 & 0 & 0 & 0 & 1 & x & x & \ldots & x \end{array}\right) \]

Partial overlap of incidence matrix rows.
\[ 
\left(
\begin{array}{ccccccccccccccccccccccccccccccccccc}
x & x & \ldots & x & 0 & 0      & 1 & 0 & 0 & 0 & x      & x & \ldots & x \\
  &   &        & x & x & \ldots & x & 0 & 0 & 0 & 0 & 0 & 1 & x & x & \ldots & x \end{array}\right) \]

The solution to our example would then look like

\[ 
\begin{array}{c}
v_1 \\ v_2 \\ v_3 \\ v_4
\end{array}
\left(
\begin{array}{cccccccccccccccccccccccccccccccccc}
x & x & \ldots & x & 001000 & x & x & \ldots & x\\
x & x & \ldots & x & 000010 & x & x & \ldots & x\\
x & x & \ldots & x & 000001 & x & x & \ldots & x\\
  &   &        &   &        &   &   & x      & x & \ldots & x & 001011 & x & x & \ldots & x\\ \end{array}\right) \]

And we could recover the number of colors from the number of groups is the string alignment (the number
of groups is recovered from the span of the solution).  Because
these flanking strings force grouping, we call them self-aligning strings.  Now we procede to describe
what these flanking regions (self-aligning strings) look like.  To motivate the process, we present
an example.  In the
following set of strings, it is obvious that the first string can not be shifted by any amount to the
right and not cause any collisions.  The first four characters will always collide with the other strings.

\[ 
\begin{array}{c}
1 \\ 2 \\ 3 \\ 4
\end{array}
\left(
\begin{array}{cccccccccccccccccccccccccccccccccccccccccccccccccccccccc}
1111 & 1000 & 1000 & 1000 & 1000 & 1000 \\
0100 & 0100 & 0100 & 0100 & 0100 & 0100 \\
0010 & 0010 & 0010 & 0010 & 0010 & 0010 \\
0001 & 0001 & 0001 & 0001 & 0001 & 0001 \\
\end{array}\right) \]

Thus, a first attempt at the self-aligning strings would be $n$ consecutive 1's followed by repeated
identity matrices.

\[ 
\begin{array}{c}
1 \\ 2 \\ 3 \\ 4
\end{array}
\left(
\begin{array}{cccccccccccccccccccccccccccccccccccccccccccccccccccccccc}
1111 & 0000 & 0000 & 0000 & 1000 & 1000 & 1000 & 1000 & 1000 & \ldots\\
0000 & 1111 & 0000 & 0000 & 0100 & 0100 & 0100 & 0100 & 0100 & \ldots\\
0000 & 0000 & 1111 & 0000 & 0010 & 0010 & 0010 & 0010 & 0010 & \ldots\\
0000 & 0000 & 0000 & 1111 & 0001 & 0001 & 0001 & 0001 & 0001 & \ldots\\
\end{array}\right) \]

The strings are grouped for clarity.  In the first 4 blocks, each row gets a sequence of 4 consecutive
1's.  The rest of the blocks are identity matrices.  In the region with indentity matrix, any submatrix
of 4 consecutive columns is a permutation matrix (ie each row has a 1 in it).  Thus, once the
4-consecutive 1's are shifted into this region, they will always collide with every string.

However, we see these are not self-aligning strings, since the consecutive 1 blocks must be shifted by as
much as $16$ places before they are in the indentity matrix region of the other strings.  For example

\[ 
\begin{array}{c}
1 \\ 2 \\ 3 \\ 4
\end{array}
\left(
\begin{array}{cccccccccccccccccccccccccccccccccccccccccccccccccccccccc}
     &      &      &      & 1111 & 0000 & 0000 & 0000 & 1000 & 1000 & 1000 & 1000 & 1000 & \ldots\\
0000 & 1111 & 0000 & 0000 & 0100 & 0100 & 0100 & 0100 & 0100 & \ldots\\
0000 & 0000 & 1111 & 0000 & 0010 & 0010 & 0010 & 0010 & 0010 & \ldots\\
0000 & 0000 & 0000 & 1111 & 0001 & 0001 & 0001 & 0001 & 0001 & \ldots\\
\end{array}\right) \]

To prevent shifts of $1$ to $15$ we can add the following types of strings to the end of the above
strings.

\[ 
\begin{array}{c}
1 \\ 2 \\ 3 \\ 4
\end{array}
\left(
\begin{array}{cccccccccccccccccccccccccccccccccccccccccccccccccccccccc}
1000 & 0000 & 0000 & 0000\\
0111 & 1111 & 1111 & 1111\\
0000 & 0000 & 0000 & 0000\\
0000 & 0000 & 0000 & 0000\\
\end{array}\right) \]

This matrix prevents the first row from shifting an amount $1$ to $15$ with the second row.  We
concatenate strings like these for every pair of rows ($4^2 = 16$).  The final idea in this
construction is that there is no limit on the number of identity matrices we included.  Thus we can
make these strings as long as we need, until the allowed overlap is a small enough fraction  (for
example, add $n^{100}$ identity matrices).  For a more precise explanation of self-aligning strings,
consult the Appendix.

\begin{theorem}
{\sc Vertex Color} ${\leq}_p$ {\sc String Pack}
\end{theorem}
\begin{proof}
Given an instance of {\sc Vertex Color}, create a {\sc String Pack} instance with the vertex-edge
incidence matrix flanked by self-aligning strings.  The number of groups in the solution to {\sc
String Pack} is the number of colors in an optimal coloring.
\end{proof}

\begin{theorem}
{\sc String Pack} is hard to approximate (No constant factor approximation).
\end{theorem}
\begin{proof}
We can approximate {\sc Vertex Color} with {\sc String Pack}.  The approximation depends on the length
of the flanking regions.  In the next section, we construct flanking strings of size $O(n^5)$.  Thus
the total size of the {\sc String Pack} instance is $O(n^6)$.  Since $n$ is the number of vertices, the
size of {\sc Vertex Color} problems are $O(m) = O(n^2)$.  So if we have an $f(n)$ approximation to {\sc
String Pack}, we get an $f(n^6) = f(m^3)$ approximation to {\sc Vertex Color}.

Since {\sc Vertex Color} is not constant factor approximable, {\sc String Pack} is not.
\end{proof}

\section{Conclusions}
\label{sec:conclusions}

In this paper, we examined the problem of accepting new video requests for pre-stored VBR 
video streams that have been pre-smoothed using any of the smoothing algorithms found in 
the literature.
We proved that this problem is an NP-complete problem by reducing it to the {\sc String Pack}
problem. 

We also presented two optimization algorithms that can be used to compute the minimum
time displacement of a new request to avoid resource overbooking. The morphology-sensitive
algorithm is capable of computing the time displacement in $O(P+n)logn$ time complexity, where
$P$ is the number of peak pairs when the first peak is selected from the schedule of the
new request, and the second from the current traffic envelope, and $n$ corresponds to the number 
of peaks in the schedule of the new video request.  

This work can be extended in several ways. In particular, when the cost to the end-user
is a variable that needs to be considered, and the cost is a function of the time displacement,
the problem can be transformed into one that finds the minimal displacement at the minimally acceptable
cost for the end-user.

Other optimization objectives could be analyzed, when given a set of requests, the requirement is
to find the displacement points of the associated schedules that produces smoothest combined schedule.

%\input{ack}

%\nocite{*}
\begin{small}
\bibliography{references}
\end{small}

\newpage
\appendix
\renewcommand{\thefigure}{A-\arabic{figure}}
\renewcommand{\thetable}{A-\arabic{table}}
\setcounter{figure}{0}
\setcounter{table}{0}
\section*{APPENDIX}

\subsection*{Self-Aligning Strings}
We now give a more precise account of Self-Aligning strings.

We call a set $S = \{s_1 \ldots s_n\}$ of strings $(n,k,L)$-aligning if the following properties hold.
\begin{enumerate}

\item $|S| = n$

\item $|s_i| = L$

\item $\{s_i(0), s_j(0)\}$ is feasible

\item $\{s_i(0), s_j(r)\}$ is not feasible for $1 \leq r \leq L-k$

\end{enumerate}

\begin{figure}[htbp]
\centerline{\psfig{figure=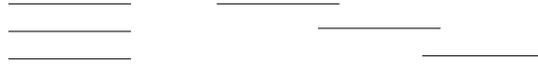,height=0.3in}}
\caption{
  Self aligning strings will either overlap completely (left) or
    overlap by some small, limited amount (right)
}
\label{fig:selfalign}
\end{figure}

For a given $(n,k,L)$, there may or may not exists a set of self-aligning strings.  We want to show a
set exists that will make the reduction in the previous section work.  That is, we need to be able to
construct them in polynomial time (it is clear from construction that it takes $O(nL)$ time to construct), and also we need
certain constraints on $n$, $k$, and $L$.  The following two constraints
are sufficient:

\begin{figure}[htbp]
\centerline{\psfig{figure=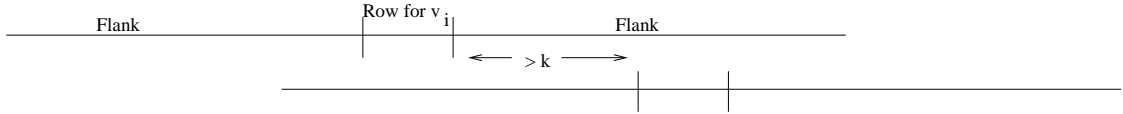,height=0.6in}}
\caption{
  The prefix copy of $a_j$ needs to overlap at least $k$ with the suffix copy of $a_i$
}
\label{fig:constraint1}
\end{figure}

First, we want the `grouping' effect.  Thus our {\sc String-Pack} strings should only be able to
overlap by at most $k$ (or equivalently only allow shifts of at least $2L + {{n}\choose{2}} - k$) .  Since our {\sc String-Pack} strings contain self-aligning strings as sub-strings; it
is obvious that shifts of $1$ to $L - k$ are not allowed.  Also, once we shift by ${{n}\choose{2}} + k
+ 1$, the prefix flanker of the shifted string overlaps the suffix flanker of the other string.  Thus
shifts of ${{n}\choose{2}} + k + 1$ to $2L + {{n}\choose{2}} - k$ are not allowed
(Figure~\ref{fig:constraint1}).  To make these 2 ranges overlap, we need
\[
L - k \geq {{n}\choose{2}} + k + 1
\]

The second constraint is to be able to recover the number of groups from the span of the solution.
Since two strings will overlap completely only if their corresponding vertices are non-adjacent, we can
recover a coloring by grouping strings that overlap completely.  Say the answer to {\sc String-Pack}
has $C$ groups of strings that overlap completely.  Since (from above) each can overlap at most $k$,
this means the span of the solution is in the range
\[
C*(|s|-k) + k \leq \mbox{span} \leq C*|s|
\]
If the answer had $C-1$ groups, the range would be
\[
(C-1)*(|s|-k) + k \leq \mbox{span} \leq (C-1)*|s|
\]
To be able to distinguish the number of completely overlapping groups from the span, we would need
\[
C*(|s|-k) + k > (C-1)*|s|
\]
That is, the smallest span from $C$ groups is larger than the largest span from $C-1$ groups.  This
yields
\[
|s| + k > k*C
\]
Since $C \leq n$, this inequality is achieved if
\[
|s| + k > kn
\]
\[
2L + {{n}\choose{2}} + k > kn
\]
\[
L > \frac{1}{2} k(n-1) - {{n}\choose{2}}
\]

These constraints are easy to achieve with the outlined construction.  To be precise, our self-aligning
strings are the rows of
\[
[R_1 R_2 \ldots R_n I^{l} P_{1,1} \ldots P_{n,n}]
\]

Where $R_i$ is the $n$x$n$ matrix
\[
\begin{array}{c}
1 \\ 2 \\ \vdots \\ i \\ \vdots \\ n
\end{array}
\left(
\begin{array}{cccccccccccccccccccccccccccccccccccccccccccccccccccccccc}
0000 \ldots 0 \\
0000 \ldots 0 \\
\vdots \\
1111 \ldots 1 \\
\vdots \\
0000 \ldots 0
\end{array}\right) \]

and $P_{i,j}$ is the $n$x$n^2$ matrix
\[
\begin{array}{c}
1 \\ 2 \\ \vdots \\ i \\ \vdots \\ \\ \vdots \\ j \\ \vdots \\ n
\end{array}
\left(
\begin{array}{cccccccccccccccccccccccccccccccccccccccccccccccccccccccc}
0000 \ldots 0 \\
0000 \ldots 0 \\
\vdots \\
1000 \ldots 0 \\
\vdots \\
0000 \ldots 0 \\
\vdots \\
0111 \ldots 1 \\
\vdots \\
0000 \ldots 0
\end{array}\right) \]

The $R_i$ give each row $n$ consecutive $1s$.  Once a string is shifted right by $n^2$, its $R_i$ will
all lie in the $I$ region of the other strings (That is, it will have $n$ consecutive 1's overlapping
the $I$ matrices).  Since in this $I$ region, a string has a $1$ every $n$ positions, this shift is not
feasible (the $n$ $1$s in the shifted string must conflict with a $1$ in the other strings).  Thus
shifts of $n^2$ through $ln$ are not feasible ($n^2$ ensures that an $R_i$ is completely in the $I$
regios;  $ln$ ensures an $R_i$ isnt shifted past the $I$ region).
The $P_{i,j}$ eliminate shifts of $1$ through $n^2-1$.  This is done explicitly, as can be seen in the
construction of the $P_{i,j}$

Thus shifts of $1$ through $ln$ are not feasible.  So the maximum overlap, $k$, is bounded by
\[
k \leq L - ln = (n*|R_i| + l*|I| + n^2|P_{i,j}|) - ln = n^2 + ln + n^4 - ln
\]
\[
k \leq n^2 + n^4
\]

This is a good result since $k$ is fixed for any size $l$.  All we want is $L > max(2k +
{{n}\choose{2}}, \frac{nk}{2})$, which we get for large enough $n$ by setting $l = n^4$ which gets $L = n^2 + ln + n^4 = O(n^5)$; a polynomial length string, as desired.

\end{document}